\documentclass[onecolumn,showpacs,showkeys,preprintnumbers,amsmath,amssymb]{revtex4}

\begin{document}


\title{Analytical Study on the Sunyaev-Zeldovich Effect for Clusters of Galaxies}

\author{Satoshi Nozawa}
 \email{snozawa@josai.ac.jp}
\affiliation{
Josai Junior College, 1-1 Keyakidai, Sakado-shi, Saitama, 350-0295,
Japan}

\author{Yasuharu Kohyama}
\affiliation{
Department of Physics, Sophia University, 7-1 Kioi-cho, Chiyoda-ku,
Tokyo, 102-8554, Japan}

\date{\today}

\begin{abstract}
  Starting from a covariant formalism of the Sunyaev-Zeldovich effect for the thermal and non-thermal distributions, we derive the frequency redistribution function identical to Wright's method assuming the smallness of the photon energy (in the Thomson limit).  We also derive the redistribution function in the covariant formalism in the Thomson limit.  We show that two redistribution functions are mathematically equivalent in the Thomson limit which is fully valid for the cosmic microwave background photon energies.  We will also extend the formalism to the kinematical Sunyaev-Zeldovich effect.  With the present formalism we will clarify the situation for the discrepancy existed in the higher order terms of the kinematical Sunyaev-Zeldovich effect.
\end{abstract}



\maketitle

\section{Introduction}

  The Sunyaev-Zeldovich (SZ) effect\cite{suny72,suny80a,suny80b,suny81}, which arises from the Compton scattering of the cosmic microwave background (CMB) photons by hot electrons in clusters of galaxies (CG), provides a useful method for studies of cosmology.  For the reviews, for example, see Birkinshaw\cite{birk99} and Carlstrom, Holder and Reese\cite{carl02}.  The original SZ formula has been derived from the Kompaneets equation\cite{komp57} in the non-relativistic approximation.  However, recent X-ray observations (for example, Schmidt et al.\cite{tuck98} and Allen et al.\cite{alle02}) have revealed the existence of high-temperature CG such as $k_{B} T_{e} \simeq $20keV.  Wright\cite{wrig79} and Rephaeli and his collaborator\cite{reph95, reph97} have done pioneering work including the relativistic corrections to the SZ effect for the CG.

In the last ten years remarkable progress has been made in theoretical studies of the relativistic corrections to the SZ effects for the CG.  Stebbins\cite{steb97} generalized the Kompaneets equation.  Challinor and Lasenby\cite{chal98} and Itoh, Kohyama and Nozawa\cite{itoh98} have adopted a relativistically covariant formalism to describe the Compton scattering process and have obtained higher-order relativistic corrections to the thermal SZ effect in the form of the Fokker-Planck approximation.  Nozawa, Itoh and Kohyama\cite{noza98} have extended their method to the case where the CG is moving with a peculiar velocity with respect to the CMB and have obtained the relativistic corrections to the kinematical SZ effect.  Their results were confirmed by Challinor and Lasenby\cite{chal99} and also by Sazonov and Sunyaev\cite{sazo98a, sazo98b}.  Itoh, Nozawa and Kohyama\cite{itoh00} have also applied the covariant formalism to the polarization SZ effect\cite{suny80b,suny81}.  Itoh and his collaborators (including the present authors) have done extensive studies on the SZ effects, which include the double scattering effect\cite{itoh01}, the effect of the motion of the observer\cite{noza05}, high precision analytic fitting formulae to the direct numerical integrations\cite{noza00,itoh02} and high precision calculations\cite{itoh04,noza06}.  The importance of the relativistic corrections is also exemplified through the possibility of directly measuring the cluster temperature using purely the SZ effect\cite{hans04}.

  On the other hand, the SZ effect in the CG has been studied also for the non-thermal distributions by several groups\cite{enss00,cola03,boeh08}.  The non-thermal distribution functions, for example, the power-law distributions, have a long tail in high electron energy regions.  Therefore the relativistic corrections for the SZ effect could be more important than the thermal distribution.

  Shimon and Rephaeli\cite{shim04} have discussed on the equivalence of different formalisms to the SZ effect.  The relativistic SZ effect has been studied analytically so far in three different approaches.  The first method is the calculation of the frequency redistribution function in the electron rest frame using the scattering probability derived by Chandrasekhar\cite{chan50}.  This method was used by Wright\cite{wrig79} and extended by Rephaeli\cite{reph95}.  We call it as Wright's method in the present paper.  The second approach solves the photon transfer equation in the electron rest frame.  This approach was used by Sazonov and Sunyaev\cite{sazo98a}.  We call it the radiative transfer method.  The third approach is the relativistic generalization of the Kompaneets equation\cite{komp57}, where the relativistically covariant Boltzmann collisional equation is solved for the photon distribution function.  This approach was used by Challinor and Lasenby\cite{chal98} and Itoh, Kohyama and Nozawa\cite{itoh98}.  We call it as the covariant formalism in the present paper.  In Shimon and Rephaeli\cite{shim04} they have shown the equivalence between Wright's method and the radiative transfer method.  They also have claimed the equivalence between Wright's method and the covariant formalism.  However, no mathematical relations are shown between the redistribution function in Wright's method and the expression of the scattering probability in the covariant formalism.  Therefore their claim is incomplete.  In the present paper we will show explicitly that two approaches are mathematically equivalent.

 On the other hand, recently Boehm and J. Lavalle\cite{boeh08} also have discussed the equivalence of the different approaches for the SZ effect in the non-thermal distribution.  They have shown that the radiative transfer method is equivalent to the covariant formalism.  However, they have concluded that Wright's method is incorrect.  In the present paper we will show that their conclusion is incorrect.  We will show that Wright's method, which has been widely used in the literature, is still fully valid.

  The fourth method for the study of the SZ effect is the direct numerical integration of the rate equation of the photon spectral distortion function.  The first-order calculation in terms of the optical depth $\tau$ was done by Itoh, Kohyama and Nozawa\cite{itoh98} for $\tau \ll 1$.  The full-order calculation was done by Dolgov et al.\cite{dolg01} for $\tau \gg 1$.  The rate equation in the present formalism has a simple form.  Therefore it is more suitable for the direct numerical application.  We will present the numerical calculation elsewhere\cite{noza09}.

  The present paper is organized as follows.  In $\S$~II, we show the equivalence between Wright's method and the covariant formalism of the SZ effect for both thermal and non-thermal distributions.  We also derive the rate equations and their formal solutions for the photon distribution function and for the spectral intensity function.  In $\S$~III we extend the formalism to the kinematical SZ effect, and derive the rate equations in Wright's method.  Finally, concluding remarks are given in $\S$~IV.

\section{Sunyaev-Zeldovich Effect}

\subsection{Equivalence between Covariant Formalism and Wright's Method}

  Let us consider that both of the CG and the observer are fixed to the CMB frame.  As a reference system, we choose the system which is fixed to the CMB.  (Three frames are identical in the present case.)  In the CMB frame, the time evolution of the photon distribution function $n(\omega)$ is written as follows\cite{itoh98}:
\begin{eqnarray}
&&\hspace{-10mm}
\frac{\partial n(\omega)}{\partial t}  =  -2 \int \frac{d^{3}p}{(2\pi)^{3}} d^{3}p^{\prime} d^{3}k^{\prime} \, W \,
\left\{ n(\omega)[1 + n(\omega^{\prime})] f(E) - n(\omega^{\prime})[1 + n(\omega)] f(E^{\prime}) \right\} \, ,  
\label{eq2-1}
\end{eqnarray}
\begin{eqnarray}
&&\hspace{-10mm}
W  =  \frac{(e^{2}/4\pi)^{2} \, \bar{X} \, \delta^{4}(p+k-p^{\prime}-k^{\prime})}{2 \omega \omega^{\prime} E E^{\prime}} \, , 
\label{eq2-2} \\
&&\hspace{-10mm}
\bar{X}  =  - \left( \frac{\kappa}{\kappa^{\prime}} + \frac{\kappa^{\prime}}{\kappa} \right) + 4 m^{4} \left( \frac{1}{\kappa} + \frac{1}{\kappa^{\prime}} \right)^{2} 
 - 4 m^{2} \left( \frac{1}{\kappa} + \frac{1}{\kappa^{\prime}} \right) \, , 
\label{eq2-3} \\
&&\hspace{-10mm}
\kappa  =  - 2 (p \cdot k) \, = \, - 2 \omega E \left( 1 - \beta \mu \right) \, ,  \label{eq2-4} \\
&&\hspace{-10mm}
\kappa^{\prime}  =   2 (p \cdot k^{\prime}) \, = \, 2 \omega^{\prime} E \left( 1 - \beta \mu^{\prime} \right) \, ,
\label{eq2-5}
\end{eqnarray}
where $e$ is the electric charge, $m$ is the electron rest mass, $W$ is the transition probability of the Compton scattering, and $f(E)$ is the electron distribution function.  The four-momenta of the initial electron and photon are $p = (E, \vec{p})$ and $k = (\omega, \vec{k})$, respectively.  The four-momenta of the final electron and photon are $p^{\prime} = (E^{\prime}, \vec{p} \, ^{\prime})$ and $k^{\prime} = (\omega^{\prime}, \vec{k}^{\prime})$, respectively.  In Eqs.~(\ref{eq2-4}) and (\ref{eq2-5}), $\beta=|\vec{p}|/E$, $\mu={\rm cos}\theta$ is the cosine between $\vec{p}$ and $\vec{k}$, and $\mu^{\prime} = {\rm cos}\theta^{\prime}$ is the cosine between $\vec{p}$ and $\vec{k}^{\prime}$.  Throughout this paper, we use the natural unit $\hbar = c = 1$, unless otherwise stated explicitly.  For later convenience we rewrite Eq.~(\ref{eq2-3}) as follows:
\begin{eqnarray}
&&\hspace{-10mm}
\bar{X}  =  \bar{X}_{A} + \bar{X}_{B}  \, ,  
\label{eq2-6}  \\
&&\hspace{-10mm}
\bar{X}_{A}  =  2 
+4m^4\left(\frac{1}{\kappa}+\frac{1}{\kappa^{\prime}}\right)^2
-4m^2\left(\frac{1}{\kappa}+\frac{1}{\kappa^{\prime}}\right) \, ,
\label{eq2-7} \\
&&\hspace{-10mm}
\bar{X}_{B}  =  -4\frac{(k\cdot k^{\prime})^2}{\kappa\kappa^{\prime}}  \, .
\label{eq2-8}
\end{eqnarray}
By eliminating the $\delta$-function, Eq.~(\ref{eq2-1}) is rewritten as follows:
\begin{eqnarray}
&&\hspace{-10mm}
\frac{\partial n(\omega)}{\partial \tau}  =  -\frac{3}{64\pi^2}
\int{d^3p}\int d\Omega_{k^{\prime}}\frac{1}{\gamma^2}\frac{1}{1-\beta\mu}
\left(\frac{\omega^{\prime}}{\omega}\right)^2\bar{X}
\nonumber \\
&&\hspace{+10mm}
\times \left\{n(\omega)[1+n(\omega^{\prime})]p_e(E)
-n(\omega^{\prime})[1+n(\omega)]p_e(E^{\prime})\right\}  \, ,
\label{eq2-9}
\end{eqnarray}
\begin{eqnarray}
&&\hspace{-10mm}
d\tau  =  n_e\sigma_T dt \, , 
\label{eq2-10}  \\
&&\hspace{-10mm}
\gamma  =  \frac{1}{\sqrt{1 - \beta^{2}}}  \, ,
\label{eq2-11}  \\
&&\hspace{-10mm}
f(E)  =  n_e \pi^2 p_e(E)  \, ,
\label{eq2-12}
\end{eqnarray}
where $n_{e}$ is the electron number density, $\sigma_{T}$ is the Thomson scattering cross section, and $p_{e}(E)$ is normalized by $\int_0^{\infty}dp p^2 p_e(E)  =  1.$  By choosing the direction of the initial electron momentum ($\vec{p}$) along $z$-axis, the photon momenta $\vec{k}$ and $\vec{k}^{\prime}$ are expressed by
\begin{eqnarray}
&&\hspace{-10mm}
\vec{k}  =  \omega \left( \sqrt{1-\mu^{2}} {\rm cos}\phi_{k}, \sqrt{1-\mu^{2}} {\rm sin}\phi_{k}, \mu \right)  \, , 
\label{eq2-13}  \\
&&\hspace{-10mm}
\vec{k}^{\prime}  = \omega^{\prime} \left( \sqrt{1-\mu^{\prime 2}} {\rm cos}\phi_{k^\prime}, \sqrt{1-\mu^{\prime 2}} {\rm sin}\phi_{k^\prime}, \mu^{\prime} \right) \, ,
\label{eq2-14}
\end{eqnarray}
where $\phi_{k}$ and $\phi_{k^\prime}$ are the azimuthal angles of $\vec{k}$ and $\vec{k}^{\prime}$, respectively.  Inserting Eqs.~(\ref{eq2-13}) and (\ref{eq2-14}) into Eqs.~(\ref{eq2-7}) and (\ref{eq2-8}), one obtains
\begin{eqnarray}
&&\hspace{-10mm}
\bar{X}_{A}  =  2
 + \frac{(1-{\rm cos} \Theta)^2}
{\gamma^4(1-\beta\mu)^2(1-\beta\mu^{\prime})^2}
 -2\frac{1-{\rm cos}\Theta}
{\gamma^2(1-\beta\mu)(1-\beta\mu^{\prime})} \, ,
\label{eq2-15} \\
&&\hspace{-10mm}
\bar{X}_{B}  =  \left(\frac{\omega}{\gamma m}\right)^2\frac{\omega^{\prime}}{\omega} \frac{(1-{\rm cos}\Theta)^2} {(1-\beta\mu)(1-\beta\mu^{\prime})} \, ,
\label{eq2-16}  \\
&&\hspace{-10mm}
\frac{\omega^{\prime}}{\omega}  =  \frac{1-\beta\mu}
{1-\beta\mu^{\prime}+(\omega/\gamma m)(1-{\rm cos}\Theta)}  \, ,
\label{eq2-17} \\
&&\hspace{-10mm}
{\rm cos}\Theta  \equiv  \mu\mu^{\prime}+\sqrt{1-\mu^2} \sqrt{1-\mu^{\prime 2}}\cos(\phi_{k}-\phi_{k^{\prime}})  \, ,
\label{eq2-18}
\end{eqnarray}
where cos$\Theta$ is the cosine between $\vec{k}$ and $\vec{k}^{\prime}$.  It should be noted that $\bar{X}_{A}$ and $\bar{X}_{B}$ will not be mixed each other under an arbitrary Lorentz transformation, because $\bar{X}_{A}$ depends only on $\mu$, $\mu^{\prime}$, $\beta$ and $\gamma$, whereas $\bar{X}_{B}$ depends also on $\omega$ and $\omega^{\prime}$.

Now let us introduce the transformations for $\mu$ and $\mu^{\prime}$ which will play a key role in the present paper.
\begin{eqnarray}
&&\hspace{-10mm}
\mu = \frac{-\mu_0+\beta}{1-\beta\mu_0} \, ,
\label{eq2-19}  \\
&&\hspace{-10mm}
 \mu^{\prime} = \frac{-\mu_0^{\prime}+\beta}{1-\beta\mu_0^{\prime}} \, ,
\label{eq2-20}
\end{eqnarray}
where $\mu_{0}={\rm cos}\theta_{0}$ and $\mu_{0}^{\prime}={\rm cos}\theta_{0}^{\prime}$ are cosines in the electron rest frame.  The suffix $0$ denotes the electron rest frame throughout this paper, unless otherwise stated explicitly.  Equations ~(\ref{eq2-19}) and (\ref{eq2-20}) are the composition of the Lorentz transformation for the photon angles from the CMB frame to the electron rest frame and the transformation $\theta_{0} \rightarrow \pi - \theta_{0}$, $\theta_{0}^{\prime} \rightarrow \pi - \theta_{0}^{\prime}$.  Note that the latter transformation is not essential.  Applying Eqs.~(\ref{eq2-19}) and (\ref{eq2-20}) to Eqs.~(\ref{eq2-15}) -- (\ref{eq2-18}), one obtains as follows:
\begin{eqnarray}
&&\hspace{-10mm}
\bar{X}_{A} = 1 + {\rm cos}^{2}\Theta_0  \, ,
\label{eq2-21} \\
&&\hspace{-10mm}
\bar{X}_{B} = \left(\frac{\omega}{\gamma m}\right)^2
\frac{\omega^{\prime}}{\omega} \frac{(1-{\rm cos}\Theta_0)^2}
{(1-\beta\mu_0)(1-\beta\mu_0^{\prime})}  \, ,
\label{eq2-22}  \\
&&\hspace{-10mm}
\frac{\omega^{\prime}}{\omega} = \frac{1-\beta\mu_0^{\prime}}
{1-\beta\mu_0+(\omega/\gamma m)(1-{\rm cos}\Theta_0)}  \, ,
\label{eq2-23} \\
&&\hspace{-10mm}
{\rm cos}\Theta_0 \equiv \mu_0\mu_0^{\prime}+\sqrt{1-\mu_0^2}
\sqrt{1-\mu_0^{\prime 2}}\cos(\phi_{k}-\phi_{k^{\prime}})  \, ,
\label{eq2-24}
\end{eqnarray}
where cos$\Theta_{0}$ is the cosine between $\vec{k}$ and $\vec{k}^{\prime}$ in the electron rest frame.  It can be seen that Eq.~(\ref{eq2-21}) was surprisingly simplified compared with Eq.~(\ref{eq2-15}).  On the other hand Eq.~(\ref{eq2-22}) did not change its form compared with Eq.~(\ref{eq2-16}).  As will see later in this section, Eqs.~(\ref{eq2-21}) and (\ref{eq2-22}) are the key points for connecting the covariant formalism with Wright's method.  The terms $\bar{X}_{A}$ and $\bar{X}_{B}$ did not mix each other by the above reason.  Furthermore $\bar{X}_{A}$ is the expression in the electron rest frame, whereas $\bar{X}_{B}$ is not, because it contains $\beta$ and $\gamma m$.

The phase space volumes are transformed as follows:
\begin{eqnarray}
&&\hspace{-10mm}
d^3p = \frac{1}{\gamma^2(1-\beta\mu_0)^2}d^3p_0  \, ,
\label{eq2-25} \\
&&\hspace{-10mm}
d\Omega_{k^{\prime}} = \frac{1}{\gamma^2(1-\beta\mu_0^{\prime})^2}d\Omega_{k^{\prime}0}  \, ,
\label{eq2-26}
\end{eqnarray}
where $d^3p_0 = p^2dpd\mu_0d\phi_p$, $d\Omega_{k^{\prime}0} = d\mu_0^{\prime}d\phi_{k^{\prime}}$.  Note that the $z$-axis was chosen along $\vec{k}$ direction for the $d^{3}p$ integration.  With these variables Eq.~(\ref{eq2-9}) is re-expressed by
\begin{eqnarray}
&&\hspace{-10mm}
\frac{\partial n(\omega)}{\partial \tau}
 = -\frac{3}{64\pi^2}
\int{d^3p_0}\int d\Omega_{k^{\prime}0}\frac{1}{\gamma^4}
\frac{1}{1-\beta\mu_0}
\frac{1}{(1-\beta\mu_0^{\prime})^2}
\left(\frac{\omega^{\prime}}{\omega}\right)^2\bar{X}
\nonumber \\
&&\hspace{+10mm}
\times
\left\{n(\omega)[1+n(\omega^{\prime})]p_e(E)
-n(\omega^{\prime})[1+n(\omega)]p_e(E^\prime)\right\}  \, .
\label{eq2-27}
\end{eqnarray}
In deriving Eq.~(\ref{eq2-27}) we used the relation $\gamma^2(1-\beta\mu) = (1-\beta\mu_0)^{-1}$.

Before proceed to the next step, some explanations might be necessary for Eq.~(\ref{eq2-27}).  In Eq.~(\ref{eq2-27}) photon zenith angles ($\mu_{0}$ and $\mu_{0}^{\prime}$) are described in the electron rest frame with the transformations of Eq.~(\ref{eq2-19}) and (\ref{eq2-20}).  On the other hand, energies ($\omega$, $\omega^{\prime}$ and $p$) and azimuthal angles ($\phi_{k}$, $\phi_{k^{\prime}}$ and $\phi_{p}$) are left in the CMB frame.  As seen later in this section, this peculiar hybrid coordinate system makes the connection from the covariant formalism to Wright's method in a straightforward manner.  It is needless to say that the familiar Klein-Nishina formula in the electron rest frame will be obtained by the Lorentz transformations $\omega = \omega_{0} \gamma \left( 1 - \beta \mu_{0} \right)$ and $\omega^{\prime} = \omega_{0}^{\prime} \gamma \left( 1 - \beta \mu_{0}^{\prime} \right)$ and inserting into Eqs.~(\ref{eq2-21}) and (\ref{eq2-22}).

  Now let us introduce an assumption which was also used in Boehm and Lavalle\cite{boeh08}.
\begin{eqnarray}
&&\hspace{-10mm}
\gamma \frac{\omega}{m} \ll 1  \, .
\label{eq2-28}
\end{eqnarray}
For the CMB ($k_{B}T_{CMB}=2.348 \times 10^{-4}$eV) photons $\omega < 5 \times 10^{-3}$eV is well satisfied.  Then $\omega/m < 1 \times 10^{-8}$, which implies $\gamma \ll 10^{8}$.  Therefore as far as the CMB photon energies are concerned, Eq.~(\ref{eq2-28}) is fully valid from the non-relativistic region to the extreme-relativistic region for the electron energies.  With Eq.~(\ref{eq2-28}) the following approximations are valid.
\begin{eqnarray}
&&\hspace{-10mm}
\frac{\omega^{\prime}}{\omega} \approx \frac{1-\beta\mu_0^{\prime}}{1-\beta\mu_0}  \, ,
\label{eq2-29}  \\
&&\hspace{-10mm}
\bar{X}_{B} = O\left[ \left(\gamma \frac{\omega}{m}\right)^2\right]  \, , 
\label{eq2-30}   \\
&&\hspace{-10mm}
E^{\prime} = E \left[ 1 + O\left(\beta \gamma \frac{\omega}{m}\right) \right] \, ,  
\label{eq2-31}  \\
&&\hspace{-10mm}
p_{e}(E^{\prime}) = p_{e}(E) \left\{
       \begin{array}{rr}
         \left[ 1 + O(T_{CMB}/T_{e}) \right]  & \hspace{0.1cm} {\rm for \, \, thermal \, \, distribution}  \\
         \left[ 1 + O(\gamma \omega/m)  \right] & \hspace{0.1cm} {\rm for \, \, power \, \, law \, \, distribution}
               \end{array}
                      \right.  \, .
\label{eq2-32}
\end{eqnarray}
As seen from Eqs.~(\ref{eq2-29})--(\ref{eq2-32}), the Thomson limit is realized in the scattering kinematics by the assumption of Eq.~(\ref{eq2-28}).  With these approximations Eq.~(\ref{eq2-27}) is reduced to
\begin{eqnarray}
&&\hspace{-10mm}
\frac{\partial n(\omega)}{\partial \tau}
 = \frac{3}{64\pi^2}
\int{d^3p_0}p_e(E)\int d\Omega_{k^{\prime}0}\frac{1}{\gamma^4}
\frac{1}{(1-\beta\mu_0)^3}(1 + {\rm cos}^{2} \Theta_0)
\left[n(\omega^{\prime})-n(\omega)\right]  \, .
\label{eq2-33}
\end{eqnarray}
Furthermore the $\phi_{k^{\prime}}$-integral can be performed and one obtains
\begin{eqnarray}
&&\hspace{-10mm}
\frac{1}{2\pi}\int_0^{2\pi}d\phi_{k^{\prime}}\left(1+{\rm cos}^{2} \Theta_0\right)
 = 1 + \mu_0^2\mu_0^{\prime 2}+\frac{1}{2}(1-\mu_0^2)(1-\mu_0^{\prime 2}) \, .
\label{eq2-34}
\end{eqnarray}
Inserting Eq.~(\ref{eq2-34}) into Eq.~(\ref{eq2-33}) and assuming the spherical symmetry for $p_{e}(E)$, one obtains as follows:
\begin{eqnarray}
&&\hspace{-10mm}
\frac{\partial n(\omega)}{\partial \tau}
 = 
\int_0^{\infty}dp p^2 p_e(E)
\nonumber \\
&&\hspace{+3mm}
\times
\int_{-1}^{1}d\mu_0 \int_{-1}^{1}d\mu_0^{\prime}\frac{1}{2\gamma^4}
\frac{1}{(1-\beta\mu_0)^3}f(\mu_0,\mu_0^{\prime})
\left[n(\omega^{\prime})-n(\omega)\right]  \, ,
\label{eq2-35}
\end{eqnarray}
\begin{eqnarray}
&&\hspace{-10mm}
f(\mu_0,\mu_0^{\prime}) = \frac{3}{8}\left[
  1 + \mu_0^2\mu_0^{\prime 2}+\frac{1}{2}(1-\mu_0^2)(1-\mu_0^{\prime 2})
\right]  \, .
\label{eq2-36}
\end{eqnarray}

According to Wright\cite{wrig79} we introduce a new variable $s$ by
\begin{eqnarray}
&&\hspace{-10mm}
e^s = \frac{\omega^{\prime}}{\omega} = \frac{1-\beta\mu_0^{\prime}}{1-\beta\mu_0}  \, ,
\label{eq2-37}
\end{eqnarray}
which implies $d\mu_0^{\prime} = -(1/\beta)(1-\beta\mu_0)e^{s}ds$.  Then Eq.~(\ref{eq2-35}) is finally rewritten by
\begin{eqnarray}
&&\hspace{-10mm}
\frac{\partial n(\omega)}{\partial \tau}
 = \int_0^{\infty}dp p^2 p_e(E) \int_{-s_{max}}^{s_{max}}ds P(s,\beta)
\left[n(e^s\omega)- n(\omega)\right]  \, ,
\label{eq2-38}  \\
&&\hspace{-10mm}
P(s,\beta)
= \frac{e^{s}}{2\beta\gamma^4}
\int_{\mu_1(s)}^{\mu_2(s)}d\mu_0(1-\beta\mu_0)
\frac{1}{(1-\beta\mu_0)^3}
f\left(\mu_0, \mu_0^{\prime} \right)   \, ,
\label{eq2-39}
\end{eqnarray}
where
\begin{eqnarray}
&&\hspace{-10mm}
s_{max} = \ln[(1+\beta)/(1-\beta)]  \, ,
\label{eq2-40} \\
&&\hspace{-10mm}
\mu_{0}^{\prime} = [1-e^s(1-\beta\mu_0)]/\beta  \, ,
\label{eq2-41}  \\
&&\hspace{-10mm}
\mu_1(s) = \left\{
\begin{array}{ll}
-1 &\quad  {\rm for} \, \, \, s \leq 0 \\
{[1-e^{-s}(1+\beta)]/\beta} &\quad {\rm for} \, \, \, s > 0
\end{array}
\right.  \, ,
\label{eq2-42} \\
&&\hspace{-10mm}
\mu_2(s) = \left\{
\begin{array}{ll}
{[1-e^{-s}(1-\beta)]/\beta} &\quad {\rm for} \, \, \, s < 0 \\
1 &\quad  {\rm for} \, \, \, s \geq 0 
\end{array}
\right. \, .
\label{eq2-43}
\end{eqnarray}
Equation (\ref{eq2-39}) is the probability for a single scattering of a photon of a frequency shift $s$ by an electron with a velocity $\beta$, which is described in the electron rest frame.  By using the identity relation $1-\beta\mu_0^{\prime} = e^{s}(1-\beta\mu_0)$, Eq.~(\ref{eq2-39}) is identical to $P(s; \beta)$ (Eq.~(7)) in Wright\cite{wrig79}.  Thus Wright's redistribution function has been derived from the covariant formalism.

  Now we will derive the redistribution function in the covariant formalism under the assumption of Eq.~(\ref{eq2-28}) (the Thomson limit).  The derivation is straightforward but lengthy.  We will give the derivation in Appendix A and will quote the result here.  The expressions which correspond to Eqs.~(\ref{eq2-38}) and (\ref{eq2-39}) in the covariant formalism (in the CMB frame) are
\begin{eqnarray}
&&\hspace{-10mm}
\frac{\partial n(\omega)}{\partial \tau}
 = \int_0^{\infty}dp p^2 p_e(E) \int_{-s_{max}}^{s_{max}}ds \tilde{P}(s,\beta)
\left[n(e^s\omega)- n(\omega)\right]  \, ,
\label{eq2-44}  \\
&&\hspace{+10mm}
\tilde{P}(s,\beta)
= \frac{e^{2s}}{2\beta\gamma^2}
\int_{\mu_1(s)}^{\mu_2(s)}d\mu^{\prime}
\tilde{f} \left(\mu, \mu^{\prime} \right)   \, ,
\label{eq2-45}  \\
&&\hspace{+10mm}
\mu = [1-e^s(1-\beta\mu^{\prime})]/\beta  \, ,
\label{eq2-46}  \\
&&\hspace{-10mm}
 \tilde{f}(\mu,\mu^{\prime})
 = \frac{3}{8}\left[2 +\frac{\displaystyle{(1-\mu\mu^{\prime})^2
+ \frac{1}{2}(1-\mu^2)(1-\mu^{\prime 2})}}
{\gamma^4(1-\beta\mu)^2(1-\beta\mu^{\prime})^2}
-2\frac{1-\mu\mu^{\prime}}
{\gamma^2(1-\beta\mu)(1-\beta\mu^{\prime})}
\right]  \, ,
\label{eq2-47}
\end{eqnarray}
where $s_{max}$, $\mu_{1}(s)$ and $\mu_{2}(s)$ are defined in Eqs.~(\ref{eq2-40}), (\ref{eq2-42}) and (\ref{eq2-43}), respectively.  In the present paragraph we show that $\tilde{P}(s, \beta)$ is identical to $P(s,\beta)$.  In order to show the equivalence, we apply the transformations of Eqs.~(\ref{eq2-19}) and (\ref{eq2-20}) to Eq.~(\ref{eq2-45}).  First, inserting Eqs.~(\ref{eq2-19}) and (\ref{eq2-20}) into Eq.~(\ref{eq2-47}), one obtains
\begin{eqnarray}
&&\hspace{+10mm}
 \tilde{f}(\mu,\mu^{\prime}) = f(\mu_{0},\mu_{0}^{\prime})  \, .
\label{eq2-48}
\end{eqnarray}
The variables $\mu^{\prime}$ and $\mu_{0}$ have the relation
\begin{eqnarray}
&&\hspace{-10mm}
\mu^{\prime} = \frac{1}{\beta} \left[1 - \frac{e^{-s}}{\gamma^{2}(1-\beta \mu_{0})} \right]  \, ,
\label{eq2-49}
\end{eqnarray}
which implies
\begin{eqnarray}
&&\hspace{-10mm}
d\mu^{\prime} = - \frac{e^{-s}}{\gamma^{2}(1-\beta \mu_{0})^{2}} d \mu_{0}
\label{eq2-50}
\end{eqnarray}
and boudary values
\begin{eqnarray}
&&\hspace{-10mm}
\mu_0 = \left\{
\begin{array}{ll}
\mu_{2}(s) &\quad {\rm at} \, \, \, \mu^{\prime} = \mu_{1}(s) \\
\mu_{1}(s) &\quad {\rm at} \, \, \, \mu^{\prime} = \mu_{2}(s) 
\end{array}
\right. \, .
\label{eq2-51}
\end{eqnarray}
Inserting Eqs.~(\ref{eq2-48})--(\ref{eq2-51}) into Eq.~(\ref{eq2-45}), one finally obtains
\begin{eqnarray}
&&\hspace{-10mm}
\tilde{P}(s,\beta)
= \frac{e^{s}}{2\beta\gamma^4}
\int_{\mu_1(s)}^{\mu_2(s)}d\mu_{0} \frac{1}{\left(1-\beta\mu_{0}\right)^{2}}
f \left(\mu_{0}, \mu_{0}^{\prime} \right)   \, ,
\label{eq2-52}
\end{eqnarray}
which is identical to Eq.~(\ref{eq2-39}).  Therefore one obtains
\begin{eqnarray}
&&\hspace{-10mm}
\tilde{P}(s,\beta) = P(s,\beta)  \, .
\label{eq2-53}
\end{eqnarray}
Thus the equivalence between the covariant formalism of the Boltzmann collisional equation\cite{itoh98} and Wright's method\cite{wrig79, reph95} has been shown mathematically under an assumption $\gamma \omega/m \ll 1$, where the assumption is fully valid for the CMB photon energies.  It should be emphasized that no non-relativistic approximations are made for the electron energies in deriving Eqs.~(\ref{eq2-39}) and (\ref{eq2-45}).  This is the reason why the calculations by two different formalisms produced same results for the SZ effect even in the relativistic electron energies.  In Appendix B, we have also shown the derivation of Eq.~(\ref{eq2-27}) in terms of the Klein-Nishina cross section formula.

  Boehm and Lavalle\cite{boeh08} also discussed the equivalence between the radiative transfer approach and the covariant formalism.  However, they concluded that Wright's method was incorrect.  We conclude that their conclusion is incorrect.  The reason why they lead the erroneous conclusion is as follows.  They start with the covariant form for the squared Compton amplitude (their Eq.~(43)).  They derived the familiar Chandrasekhar's form (their Eq.~(50)) by taking the non-relativistic limit ($\beta \rightarrow 0$) in their Eq.~(49).  Because of the non-relativistic approximation they used, they concluded that Wright's method (Eq.~(50)) should not be used for the relativistic calculation.  On the other hand, we have also started with the same covariant form for the squared Compton amplitude.  We have derived the same expression (Eq.~(34)) without taking the non-relativistic limit.  We have shown that Eq.~(34) is connected to its covariant form by the Lorentz transformations of Eqs.~(19) and (20).  Therefore Wright's method is equivalent to the covariant formalism.   We conclude that their criticism is incorrect.  Shimon and Rephaeli\cite{shim04} also claimed the equivalence between the covariant formalism and Wright's method.  Their Eq.~(19) looks similar to Eq.~(\ref{eq2-38}), however, no mathematical relations are shown explicitly in their paper between $W$ in their Eq.~(19) and $P(s;\beta)$ of Wright\cite{wrig79}.

\subsection{Rate Equations and Formal Solutions}

  We now proceed to derive the rate equations and their formal solutions.  Since two formalisms are equivalent, one can use either $P(s,\beta)$ or $\tilde{P}(s,\beta)$.  We start with Eq.~(\ref{eq2-38}) and rewrite as follows:
\begin{eqnarray}
&&\hspace{-10mm}
\frac{\partial n(\omega)}{\partial \tau}
 =
\int_{-\infty}^{\infty}ds P_1(s)
\left[n(e^s\omega)- n(\omega)\right] \, ,
\label{eq2-54}   \\
 &&\hspace{-10mm}
 P_1(s) = \int_{\beta_{min}}^{1}d\beta\beta^2\gamma^5 \tilde{p}_e(\beta)P(s,\beta)  \, ,
\label{eq2-55}  \\
&&\hspace{-10mm}
\beta_{min} = (1-e^{-|s|})/(1+e^{-|s|})  \, ,
\label{eq2-56}
\end{eqnarray}
where $\tilde{p}_{e}(\beta) \equiv m^{3} p_{e}(E)$.  As seen from Eq.~(\ref{eq2-55}), $P_{1}(s)$ is the probability for a single scattering of a photon of a frequency shift $s$ averaged over the electron distribution function, which is so called the redistribution function of a shift $s$.  The total probability is $\int_{-\infty}^{\infty}ds P_1(s) = 1$.  Multiplying $\omega^{3}$ to Eq.~(\ref{eq2-54}), one obtains the rate equation for the spectral intensity function.
\begin{eqnarray}
&&\hspace{-10mm}
\frac{\partial I(\omega)}{\partial \tau}
 =
\int_{-\infty}^{\infty}ds
{P}_1(s)
\left[e^{-3s}I(e^{s}\omega)- I(\omega)\right]  \, ,
\label{eq2-57}
\end{eqnarray}
where $I(\omega)=\omega^{3} n(\omega)/2\pi^{2}$ is the spectral intensity function for $\omega$.  Now let us introduce the following key identity relations:
\begin{eqnarray}
&&\hspace{-10mm}
P(s, \beta)e^{-3s} = P(-s,\beta) \, , \, \, \,P_{1}(s)e^{-3s} = P_{1}(-s) \, .
\label{eq2-58}
\end{eqnarray}
The derivation is straightforward.  Inserting Eq.~(\ref{eq2-58}) in Eq.~(\ref{eq2-57}) and replacing $s$ by $-s$, one obtains the rate equation for the spectral intensity function.
\begin{eqnarray}
&&\hspace{-10mm}
\frac{\partial I(\omega)}{\partial \tau}
 =
\int_{-\infty}^{\infty}ds
{P}_1(s)
\left[I(e^{-s}\omega)- I(\omega)\right]  \, .
\label{eq2-59}
\end{eqnarray}
It should be remarked that $n(e^{s}\omega)$ appears in RHS of Eq.~(\ref{eq2-54}), whereas $I(e^{-s}\omega)$ appears in RHS of Eq.~(\ref{eq2-59}).  It is also straightforward to show that Eq.~(\ref{eq2-54}) satisfies the photon number conservation.
\begin{eqnarray}
&&\hspace{-10mm}
\frac{d}{d \tau} \int_{0}^{\infty} d\omega \omega^{2} n(\omega) = 0  \, .
\label{eq2-60}
\end{eqnarray}

  Let us now derive formal solutions for the rate equations Eq.~(\ref{eq2-54}) and Eq.~(\ref{eq2-59}).  We consider an ideal condition that the CG is infinitely large.  We introduce a new function $\tilde{n}(\omega, \tau)$ by
\begin{eqnarray}
&&\hspace{-10mm}
n(\omega) \equiv e^{-\tau} \tilde{n}(\omega, \tau) \, .
\label{eq2-61}
\end{eqnarray}
By inserting Eq.~(\ref{eq2-61}) into Eq.~(\ref{eq2-54}), one obtains the equation for $\tilde{n}(\omega, \tau)$.
\begin{eqnarray}
&&\hspace{-10mm}
\frac{\partial \tilde{n}(\omega, \tau)}{\partial \tau}
 =
\int_{-\infty}^{\infty}ds P_1(s) \tilde{n}(e^s\omega, \tau) \, ,
\label{eq2-62}
\end{eqnarray}
where $\int_{-\infty}^{\infty}ds P_1(s) = 1$ was used.  Equation (\ref{eq2-62}) can be integrated and one has
\begin{eqnarray}
&&\hspace{-20mm}
\tilde{n}(\omega, \tau) = n_0(\omega) +
\int_{0}^{\tau}d\lambda
\int_{-\infty}^{\infty}ds P_1(s)
\tilde{n}(e^{s}\omega, \lambda)  \, .
\label{eq2-63}
\end{eqnarray}
In deriving Eq.~(\ref{eq2-63}) an initial condition $\tilde{n}(\omega, \tau=0) = n_{0}(\omega)$ was used, where $n_{0}(\omega)$ is the initial photon distribution function.  We solve Eq.~(\ref{eq2-63}) with a successive approximation method.  The first-order term is obtained by inserting $n_{0}(\omega)$ into RHS of Eq.~(\ref{eq2-63}).
\begin{eqnarray}
&&\hspace{-20mm}
\tilde{n}_{1}(\omega, \tau) = n_0(\omega)
 + \tau \int_{-\infty}^{\infty}ds P_1(s)
n_{0}(e^{s}\omega)  \, .
\label{eq2-64}
\end{eqnarray}
The second-order term is also obtained by inserting $\tilde{n}_{1}(\omega, \tau)$ into RHS of Eq.~(\ref{eq2-63}).
\begin{eqnarray}
&&\hspace{-25mm}
\tilde{n}_2(\omega,\tau) = n_0(\omega) +
\tau \int_{-\infty}^{\infty}ds P_1(s) n_0(e^{s}\omega)
\nonumber \\
&&\hspace{+6mm}
+\frac{\tau^2}{2!}
\int_{-\infty}^{\infty}ds P_2(s) n_0(e^{s}\omega)  \, ,
\label{eq2-65} \\
&&\hspace{-10mm}
P_2(s) \equiv \int_{-\infty}^{\infty}ds_1 P_1(s_1) P_1(s-s_1) \, ,
\label{eq2-66}
\end{eqnarray}
where $P_{2}(s)$ is the probability (redistribution function) of a shift $s$ for the double scattering.  By repeating the above procedure $N+1$ times, one obtains the $(N+1)$-th order term.
\begin{eqnarray}
&&\hspace{-12mm}
\tilde{n}_{N+1}(\omega,\tau) = n_0(\omega) +
\sum_{j=1}^N\frac{\tau^j}{j!}
\int_{-\infty}^{\infty}ds P_j(s)
n_0(e^{s}\omega)  \, ,
\label{eq2-67}  \\
&&\hspace{-2mm}
P_{j}(s) = \int_{-\infty}^{\infty}ds_1 P_1(s_1)\cdots
\int_{-\infty}^{\infty}ds_{j-1} P_1(s_{j-1}) P_1(s-\sum_{i=1}^{j-1}s_i) \, ,
\label{eq2-68}
\end{eqnarray}
where $P_{j}(s)$ is the probability (redistribution function) of a shift $s$ for the multiple scattering of the $j$-th order.  By taking the limit $N\to\infty$ in Eq.~(\ref{eq2-67}) and replacing $\lim_{N\to\infty}\tilde{n}_N(\omega,\tau) = \tilde{n}(\omega,\tau)$, one finally obtains the formal solution for $n(\omega)$.
\begin{eqnarray}
&&\hspace{-10mm}
n(\omega) = e^{-\tau} n_0(\omega) +
\int_{-\infty}^{\infty}ds
P(s,\tau) n_0(e^{s}\omega)  \, ,
\label{eq2-69}  \\
&&\hspace{-10mm}
P(s,\tau) =
\sum_{j=1}^{\infty}\frac{\tau^j e^{-\tau}}{j!} P_j(s)  \, .
\label{eq2-70}
\end{eqnarray}
Multiplying $\omega^{3}$ to Eq.~(\ref{eq2-69}) and using $P(s,\tau)e^{-3s}=P(-s,\tau)$, and also replacing $s$ by $-s$, one obtains the formal solution for $I(\omega)$.
\begin{eqnarray}
&&\hspace{-10mm}
I(\omega) = e^{-\tau} I_0(\omega) + 
\int_{-\infty}^{\infty}ds
P(s,\tau) I_0(e^{-s}\omega)  \, ,
\label{eq2-71}
\end{eqnarray}
where $I_{0}(\omega)=\omega^{3}/(2\pi^{2}) n_{0}(\omega)$.  Note that this solution can be also derived directly from Eq.~(\ref{eq2-59}).  Note also that Eq.~(\ref{eq2-70}) is the Poisson distribution function.  The distribution function is commonly used, for example, in Birkinshaw\cite{birk99}.  In the present paper, however, Eq.~(\ref{eq2-70}) is derived as a natural consequence of the present formalism.

In practical cases, the CG has a finite size and the optical depth is small ($\tau \ll 1$), therefore the first order approximation is sufficiently accurate for the study of the SZ effect.  From Eqs.~(\ref{eq2-69})--(\ref{eq2-71}) one obtains the following familiar forms for the distortion functions.
\begin{eqnarray}
&&\hspace{-10mm}
\Delta n(\omega) \equiv n(\omega) - n_{0}(\omega)  \,
\nonumber  \\
&&\hspace{+3mm}
\approx \tau \int_{-\infty}^{\infty}ds
P_{1}(s) \left[ n_0(e^{s}\omega) - n_0(\omega) \right]  \, ,
\label{eq2-72}  \\
&&\hspace{-10mm}
\Delta I(\omega) \equiv I(\omega) - I_{0}(\omega)  \,
\nonumber  \\
&&\hspace{+3mm}
\approx \tau \int_{-\infty}^{\infty}ds
P_{1}(s) \left[ I_0(e^{-s}\omega) - I_0(\omega) \right]  \, ,
\label{eq2-73}  \\
&&\hspace{-1mm}
\tau = \sigma_{T} \int d \ell n_{e}  \, .
\label{eq2-74}
\end{eqnarray}
The integral in Eq.~(\ref{eq2-74}) is done over the photon path length in the CG.

\section{Kinematical Sunyaev-Zeldovich Effect}

  Let us now consider the case that the CG is moving with a peculiar velocity $\vec{\beta}_{c}$ (=$\vec{v}_{c}/c$) with respect to the CMB.  As a reference system, we choose the system which is fixed to the CMB.  The $z$-axis is fixed to a line connecting the observer and the center of mass of the CG.  (We assume that the observer is fixed to the CMB frame.)  In the present paper we choose the positive direction of the $z$-axis as the conventional one, i.e. the direction of the propagation of a photon from the observer to the cluster, which is opposite to that of Nozawa, Itoh and Kohyama\cite{noza98}.  In the CMB frame, the time evolution of the photon distribution function $n(\omega)$ is same as for the thermal SZ effect as shown in Nozawa, Itoh and Kohyama\cite{noza98}.  They are given by Eqs.~(\ref{eq2-1})--(\ref{eq2-5}).  The electron distribution functions are Lorentz invariant and are related as follows:
\begin{eqnarray}
&&\hspace{-10mm}
f(E) = f_{c}(E_{c})  \, ,
\label{eq3-1}  \\
&&\hspace{-10mm}
f(E^{\prime}) = f_{c}(E^{\prime}_{c})  \, ,
\label{eq3-2}
\end{eqnarray}
\begin{eqnarray}
&&\hspace{-10mm}
E_{c} = E \gamma_{c} \left(1 + \vec{\beta}_{c} \cdot \vec{\beta} \right)  \, ,
\label{eq3-3}  \\
&&\hspace{-10mm}
E^{\prime}_{c} = E^{\prime} \gamma_{c} \left(1+ \vec{\beta}_{c} \cdot \vec{\beta} \, ^{\prime} \right)    \, , 
\label{eq3-4}  \\
&&\hspace{-10mm}
\gamma_{c} = \frac{1}{\sqrt{1 - \beta_{c}^{2}}} \, ,
\label{eq3-5}
\end{eqnarray}
where the suffix $c$ denotes the CG frame.  Therefore the formalism of $\S$~II will be directly applicable to the present case.  A modification should be made to the electron distribution function $p_{e}(E)$ by
\begin{eqnarray}
&&\hspace{-22mm}
p_{e}(E) = p_{e, c} \left( E \gamma_{c} \left[ 1 + \vec{\beta}_{c} \cdot \vec{\beta} \right]  \right)  \, ,
\label{eq3-6}
\end{eqnarray}
where $p_{e,c}(E_{c})$ is normalized by $\int_{0}^{\infty} dp_{c} p_{c}^{2} p_{e,c}(E_{c}) = 1$.  To proceed the calculation, one expresses the product $\vec{\beta}_{c} \cdot \vec{\beta}$ in the coordinate system where $\vec{k}$ is parallel to the $z$-axis.  Then one obtains
\begin{eqnarray}
&&\hspace{-10mm}
\vec{\beta}_{c} \cdot \vec{\beta} = \beta_{c} \beta \left\{ \mu_{c}\mu + \sqrt{1-\mu_{c}^2}\sqrt{1-\mu^2} \cos(\phi_{c}-\phi_{p}) \right\} \, ,
\label{eq3-7}
\end{eqnarray}
where $\mu_{c}$ and $\phi_{c}$ are the cosine of the zenith angle and the azimuthal angle of $\vec{\beta}_{c}$, respectively.  By applying the transformation of Eq.~(\ref{eq2-19}) to Eq.~(\ref{eq3-7}), one obtains
\begin{eqnarray}
&&\hspace{-10mm}
\vec{\beta}_{c} \cdot \vec{\beta} = \frac{\beta_{c} \beta}{1-\beta\mu_0}
\left[\mu_{c}(-\mu_0+\beta)+\frac{1}{\gamma}\sqrt{1-\mu_{c}^2}\sqrt{1-\mu_0^2}
\cos(\phi_{c}-\phi_{p})\right]  \, .
\label{eq3-8}
\end{eqnarray}
Inserting Eqs.~(\ref{eq3-6}) and (\ref{eq3-8}) into Eq.~(\ref{eq2-33}), one obtains the expression for the CG with non-zero peculiar velocity in Wright's method.
\begin{eqnarray}
&&\hspace{-10mm}
\frac{\partial n(\omega)}{\partial \tau}
 = \int_0^{\infty}dp p^2
\int_{-1}^{1}d\mu_0 \int_{-1}^{1}d\mu^{\prime}_0\frac{1}{2\gamma^4}
\frac{1}{(1-\beta\mu_0)^3}f(\mu_0,\mu_0^{\prime})
\nonumber \\
&&\hspace{+5mm}
\times
\frac{1}{2\pi}\int_{0}^{2\pi}d\phi_p
 p_{e,c}\left(E \gamma_{c} \left[1+ \vec{\beta}_{c} \cdot \vec{\beta} \right]\right) \left[n(\omega^{\prime})-n(\omega)\right]  \, .
\label{eq3-9}
\end{eqnarray}

  Shimon and Rephaeli\cite{shim04} also obtained the expression for the kinematical SZ effect based upon Wright's method, which is similar to Eq.~(\ref{eq3-9}).  For the expression of $\vec{\beta}_{c} \cdot \vec{\beta}$, Eq.~(\ref{eq3-8}) agrees with their Eq.~(39).   As discussed in their paper, however, they have an extra factor $\gamma_{c} \left(1+ \vec{\beta}_{c} \cdot \vec{\beta} \right)$ in Eq.~(\ref{eq3-9}) which comes from $E_{c}/E$ in their phase space factor, see their Eq.~(37).  As discussed also in Nozawa, Itoh, Suda and Ohhata\cite{noza06}, the reason of the discrepancy is because they used the phase space in the CG frame instead of the CMB frame.  As far as the present formalism is concerned, we have used the CMB frame as a reference system.  Therefore there are no extra factors needed in Eq.~(\ref{eq3-9}).  We conclude that the result of Shimon and Rephaeli is in error by the extra factor.

  Let us now proceed with Eq.~(\ref{eq3-9}).  For most of the CG, $\beta_{c} \ll 1$ is realized.  For example, $\beta_{c} \approx$ 1/300 for a typical value of the peculiar velocity $v_{c}$=1000 km/s.  In Nozawa, Itoh and Kohyama\cite{noza98} they made an expansion in terms of $\beta_{c}$ in the Fokker-Planck approximation.  They found that $O(\beta_{c}^{2})$ terms are negligible for most of the CG.  Therefore we will keep $O(\beta_{c})$ terms and neglect higher-order terms in the present paper.  In this approximation the electron distribution function is approximated as follows:
\begin{eqnarray}
&&\hspace{-10mm}
p_{e,c}(E_{c}) \approx p_{e}(E) \left\{
\begin{array}{ll}
\left( 1 - \displaystyle{\frac{a}{\beta^{2}}} \vec{\beta}_{c} \cdot \vec{\beta}  \right) & \, \, \, \, \, {\rm for} \, \, \,  p_{e}(E) \propto p^{-a}  \\
\left( 1 - a \vec{\beta}_{c} \cdot \vec{\beta} \right) & \, \, \, \, \, {\rm for} \, \, \, p_{e}(E) \propto E^{-a}  \\
\displaystyle{ \left( 1 - \frac{E}{k_{B}T_{e}} \vec{\beta}_{c} \cdot \vec{\beta}  \right) } & \, \, \, \, \, {\rm for} \, \, \, p_{e}(E) \propto {\rm exp}(-E/k_{B}T_{e}) 
\end{array}
\right. \, .
\label{eq3-10}
\end{eqnarray}
For simplicity, we consider the thermal distribution function. (Only a minor modification will be needed for the power-law distributions.)  Inserting Eq.~(\ref{eq3-8}) into Eq.~(\ref{eq3-10}) the integral for the azimuthal angle is performed.
\begin{eqnarray}
&&\hspace{-10mm}
\frac{1}{2\pi}\int_{0}^{2\pi}d\phi_p
 p_{e,c}\left(E \gamma_{c} \left[1 + \vec{\beta}_{c} \cdot \vec{\beta} \right]\right) \approx p_{e}(E) \left[ 1 + \beta_{c} \mu_{c} \left( \frac{\gamma}{\theta_{e}} \right) \left( \frac{ \beta \mu_{0} - \beta^2}{1 - \beta \mu_{0}} \right) \right] \, ,
\label{eq3-11}
\end{eqnarray}
where $\theta_{e} \equiv k_{B}T_{e}/m$.  Repeating the same procedure done in $\S$~II, one obtains the rate equations for the case of the CG with nonzero peculiar velocity.
\begin{eqnarray}
&&\hspace{-10mm}
\frac{\partial n(\omega)}{\partial \tau}
 =
\int_{-\infty}^{\infty}ds P_1(s,\beta_{c,z})
\left[n(e^s\omega)- n(\omega)\right] \, ,
\label{eq3-12}   \\
&&\hspace{-10mm}
\frac{\partial I(\omega)}{\partial \tau}
 =
\int_{-\infty}^{\infty}ds
{P}_1(s,\beta_{c,z})
\left[e^{-3s}I(e^{s}\omega)- I(\omega)\right]  \, ,
\label{eq3-13}  \\
&&\hspace{+10mm}
P_1(s,\beta_{c,z}) = P_{1}(s) + \beta_{c,z} P_{1, K}(s)  \, ,
\label{eq3-14}
\end{eqnarray}
where $P_{1}(s)$ is Eq.~(\ref{eq2-55}) and $\beta_{c,z}=\beta_{c} \mu_{c}$ is the peculiar velocity parallel to the observer, because the photon direction is along $z$-axis.  In Eq.~(\ref{eq3-14}), $P_{1,K}(s)$ is the redistribution function due to the peculiar velocity of the CG.  It is given as
\begin{eqnarray}
&&\hspace{-10mm}
 P_{1,K}(s) = \int_{\beta_{min}}^{1}d\beta\beta^2\gamma^5 \tilde{p}_e(\beta)P_{K}(s,\beta)  \, ,
\label{eq3-15}  \\
&&\hspace{-10mm}
P_{K}(s,\beta)
= \frac{e^{s}}{2\beta\gamma^4} \left(\frac{\gamma}{\theta_{e}} \right)
\int_{\mu_1(s)}^{\mu_2(s)}d\mu_0 (\beta \mu_{0}-\beta^2)
\frac{1}{(1-\beta\mu_0)^3}
f\left(\mu_0, \mu_{0}^{\prime} \right)   \, ,
\label{eq3-16}
\end{eqnarray}
where $\mu_{0}^{\prime}$, $\mu_{1}(s)$, $\mu_{2}(s)$ and $\beta_{min}$ are defined in Eqs.~(\ref{eq2-41}), (\ref{eq2-42}), (\ref{eq2-43}) and (\ref{eq2-56}), respectively.  It should be remarked that Eq.~(\ref{eq3-13}) is expressed by $e^{-3s}I(e^{s}\omega)$ instead of $I(e^{-s}\omega)$ in Eq.~(\ref{eq2-59}).  This is because $P(s, \beta) e^{-3s}=P(-s, \beta)$ as shown in Eq.~(\ref{eq2-58}), however, $P_{K}(s, \beta)e^{-3s} \neq P_{K}(-s, \beta)$.  For the power-law distributions, $(\gamma/\theta_{e}$) should be replaced by $a/\beta^2$ and $a$ in Eq.~(\ref{eq3-16}) for the $p$-power distribution and the $E$-power distribution, respectively.

  Finally, one obtains the distortions of the photon spectrum and the spectral intensity in the first order approximation.
\begin{eqnarray}
&&\hspace{-10mm}
\Delta n(\omega)
\approx \tau \int_{-\infty}^{\infty}ds
P_{1}(s, \beta_{c,z}) \left[ n_0(e^{s}\omega) - n_0(\omega) \right]  \, ,
\label{eq3-17}  \\
&&\hspace{-10mm}
\Delta I(\omega)
\approx \tau \int_{-\infty}^{\infty}ds
P_{1}(s, \beta_{c,z}) \left[ e^{-3s}I_0(e^{s}\omega) - I_0(\omega) \right]  \, .
\end{eqnarray}

\section{Concluding Remarks}

  We started with a covariant Boltzmann collisional equation of the SZ effect shown in Itoh, Kohyama and Nozawa\cite{itoh98} for thermal and non-thermal distributions.  First we have applied a rational transformation (Eqs.~(\ref{eq2-19}) and (\ref{eq2-20})) to the photon angles, which is essentially a Lorentz transformation for photon angles from the CMB frame to the electron rest frame.  The transformation has made the expression for the transition probability surprisingly concise form.  Then we have introduced an assumption used by Boehm and Lavalle\cite{boeh08}, namely $\gamma \omega/m \ll 1$ (the Thomson limit).  The assumption is fully valid for the CMB photon energies.  Under the assumption, we have derived the redistribution function $P(s,\beta)$, which is the probability for a single scattering of a photon of a frequency shift $s$ by a electron with a velocity $\beta$.  The obtained redistribution function is identical to that of derived with Wright's method\cite{wrig79, reph95}.

  Similarly, starting from the covariant Boltzmann collisional equation of the SZ effect for thermal and non-thermal distributions, we have derived the redistribution function $\tilde{P}(s,\beta)$ in the covariant formalism under the assumption $\gamma \omega/m \ll 1$.  We have shown that $\tilde{P}(s,\beta)$ is identical to $P(s,\beta)$.  They are connected by the Lorentz transformation of Eqs.~(\ref{eq2-19}) and (\ref{eq2-20}).  Thus we have shown mathematically that Wright's method is equivalent to the covariant formalism under the assumption $\gamma \omega/m \ll 1$.  This result guarantees that existing works which used Wright's method, for example, Birkinshaw\cite{birk99}, En{\ss}lin and Kaiser\cite{enss00} and Colafrancesco et al.\cite{cola03}, are still fully valid.  This result also explains the reason why two different calculations for the thermal SZ effect agree extremely well even for the relativistic electron energies.

  We have also extended the present formalism to the kinematical SZ effect.  Starting from the covariant Boltzmann collisional equation for the kinematical SZ effect, we have repeated the same procedure.  We have derived the redistribution function for the CG with nonzero peculiar velocity in Wright's method.  We have compared the present result with that of Shimon and Rephaeli\cite{shim04}.  The obtained redistribution function is differ by a factor $\gamma_{c} \left(1+ \vec{\beta}_{c} \cdot \vec{\beta} \right)$.  We have clarified the discrepancy between their result and others\cite{noza98,sazo98a,chal99}.  Their result is in error by the factor.

\begin{acknowledgments}
We wish to acknowledge Professor N. Itoh for enlightening us on this subject and also for giving us many useful suggestions.  We would also like to thank our referee for valuable suggestions.
\end{acknowledgments}

\appendix

\section{Redistribution Function in Covariant Formalism}

  In this appendix we will derive the redistribution function in the covariant formalism.  The starting equation is Eq.~(\ref{eq2-9}).
\begin{eqnarray}
&&\hspace{-10mm}
\frac{\partial n(\omega)}{\partial \tau}  =  -\frac{3}{64\pi^2}
\int{d^3p}\int d\Omega_{k^{\prime}}\frac{1}{\gamma^2}\frac{1}{1-\beta\mu}
\left(\frac{\omega^{\prime}}{\omega}\right)^2\bar{X}
\nonumber \\
&&\hspace{+10mm}
\times \left\{n(\omega)[1+n(\omega^{\prime})]p_e(E)
-n(\omega^{\prime})[1+n(\omega)]p_e(E^{\prime})\right\}  \, .
\label{eqa-1}
\end{eqnarray}
Then we assume the Thomson limit $\gamma \omega/m \ll 1$, which implies the approximations
\begin{eqnarray}
&&\hspace{-10mm}
\frac{\omega^{\prime}}{\omega} \approx \frac{1-\beta\mu}{1-\beta\mu^{\prime}}
\label{eqa-2}
\end{eqnarray}
and $\bar{X}_{B} \ll 1$, $E^{\prime} \approx E$ and $p_{e}(E^{\prime}) \approx  p_{e}(E)$.  Under the assumption, Eq.~(\ref{eqa-1}) is approximated as
\begin{eqnarray}
&&\hspace{-10mm}
\frac{\partial n(\omega)}{\partial \tau}
 = \frac{3}{64\pi^2}
\int{d^3p}p_e(E)\int d\Omega_{k^{\prime}}\frac{1}{\gamma^2}
\frac{1-\beta\mu}{(1-\beta\mu^{\prime})^2}
\bar{X}_{A}
[n(\omega^{\prime})-n(\omega)]  \, ,
\label{eqa-3}
\end{eqnarray}
where $\bar{X}_{A}$ is given by Eq.~(\ref{eq2-15}).  In Eq.~(\ref{eqa-3}) the $\phi_{k^{\prime}}$-integration can be done as
\begin{eqnarray}
&&\hspace{-10mm}
 \frac{1}{2\pi}\int_{0}^{2\pi}\bar{X}_{A} d\phi_{k^{\prime}}
 = 2 +\frac{\displaystyle{(1-\mu\mu^{\prime})^2
+ \frac{1}{2}(1-\mu^2)(1-\mu^{\prime 2})}}
{\gamma^4(1-\beta\mu)^2(1-\beta\mu^{\prime})^2}
\nonumber \\
&&\hspace{+25mm}
-2\frac{1-\mu\mu^{\prime}}
{\gamma^2(1-\beta\mu)(1-\beta\mu^{\prime})}  \, .
\label{eqa-4}
\end{eqnarray}
Assuming the spherical symmetry for $p_{e}(E)$, Eq.~(\ref{eqa-3}) is further simplified.
\begin{eqnarray}
&&\hspace{-10mm}
\frac{\partial n(\omega)}{\partial \tau}
 =
\int_{0}^{\infty} dp p^2p_e(E)\int_{-1}^{1}d\mu\int_{-1}^{1}d\mu^{\prime}
\frac{1}{2\gamma^2}
\frac{1-\beta\mu}{(1-\beta\mu^{\prime})^2}
\tilde{f}(\mu,\mu^{\prime})[n(\omega^{\prime})-n(\omega)]  \, ,
\label{eqa-5}  \\
&&\hspace{-10mm}
 \tilde{f}(\mu,\mu^{\prime})
 = \frac{3}{8}\left[2 +\frac{\displaystyle{(1-\mu\mu^{\prime})^2
+ \frac{1}{2}(1-\mu^2)(1-\mu^{\prime 2})}}
{\gamma^4(1-\beta\mu)^2(1-\beta\mu^{\prime})^2}
-2\frac{1-\mu\mu^{\prime}}
{\gamma^2(1-\beta\mu)(1-\beta\mu^{\prime})}
\right]  \, .
\label{eqa-6}
\end{eqnarray}

  Now let us introduce a new variable $s$ by
\begin{eqnarray}
&&\hspace{-10mm}
e^s = \frac{\omega^{\prime}}{\omega} = \frac{1-\beta\mu}{1-\beta\mu^{\prime}}  \, ,
\label{eqa-7}
\end{eqnarray}
which implies $d\mu = -(1/\beta)(1-\beta\mu^{\prime})e^{s}ds$.  Then Eq.~(\ref{eqa-5}) is finally rewritten by
\begin{eqnarray}
&&\hspace{-10mm}
\frac{\partial n(\omega)}{\partial \tau}
 = \int_0^{\infty}dp p^2 p_e(E) \int_{-s_{max}}^{s_{max}}ds \tilde{P}(s,\beta)
\left[n(e^s\omega)- n(\omega)\right]  \, ,
\label{eqa-8}  \\
&&\hspace{+10mm}
\tilde{P}(s,\beta)
= \frac{e^{2s}}{2\beta\gamma^4}
\int_{\mu_1(s)}^{\mu_2(s)}d\mu^{\prime}
\tilde{f}\left(\mu, \mu^{\prime} \right)   \, ,
\label{eqa-9}  \\
&&\hspace{+10mm}
\mu = \left[1 - e^{s}(1-\beta \mu^{\prime}) \right]/\beta  \, ,
\label{eqa-10}
\end{eqnarray}
where $s_{max}$, $\mu_{1}(s)$ and $\mu_{2}(s)$ are given by Eqs.~(\ref{eq2-40}), (\ref{eq2-42}) and (\ref{eq2-43}), respectively.  Equation (\ref{eqa-9}) is the redistribution function in the covariant formalism, which is described in the CMB frame.

\section{Klein-Nishina Cross Section}

  In this appendix we will derive Eq.~(\ref{eq2-27}) in terms of familiar Klein-Nishina cross section formula.  Notations are same as those in the main text, unless otherwise stated explicitly.  As a reference frame we choose the electron rest frame.  The energy-momentum conservation gives the relation for the photon energies as follows:
\begin{eqnarray}
&&\hspace{-10mm}
\frac{\omega_0^{\prime}}{\omega_0} = \frac{1}{1+(\omega_0/m)(1-\cos\Theta_{0})} \, ,
\label{eqb-1}  \\
&&\hspace{-10mm}
{\rm cos}\Theta_0 \equiv \mu_0\mu_0^{\prime}+\sqrt{1-\mu_0^2}
\sqrt{1-\mu_0^{\prime 2}}\cos(\phi_{k_{0}}-\phi_{k_{0}^{\prime}})  \, ,
\label{eqb-2}
\end{eqnarray}
where $\Theta_{0}$ is the scattering angle.  The Klein-Nishina cross section formula in the electron rest frame is expressed by
\begin{eqnarray}
&&\hspace{-10mm}
\frac{d \sigma}{d \Omega_{k_{0}^{\prime}}} = \frac{1}{2} r_{e}^{2} \left(\frac{\omega_{0}^{\prime}}{\omega_{0}} \right)^{2} 
\left( \frac{\omega_{0}^{\prime}}{\omega_{0}} + \frac{\omega_{0}}{\omega_{0}^{\prime}} - \sin^{2} \Theta_{0} \right) \, ,
\label{eqb-3}
\end{eqnarray}
where $r_{e}$ is the classical electron radius.  With Eq.~(\ref{eqb-1}) one obtains the following usefull relation.
\begin{eqnarray}
&&\hspace{-10mm}
\frac{\omega_{0}^{\prime}}{\omega_{0}} + \frac{\omega_{0}}{\omega_{0}^{\prime}} = 2 + \left(\frac{\omega_{0}}{m} \right)^{2} \frac{\omega_{0}^{\prime}}{\omega_{0}} \left( 1 - \cos\Theta_{0} \right)^{2} \, .
\label{eqb-4}
\end{eqnarray}
Inserting Eq.~(\ref{eqb-4}) into Eq.~(\ref{eqb-3}) one can rewrite the Klein-Nishina formula as follows:
\begin{eqnarray}
&&\hspace{-10mm}
\frac{d \sigma}{d \Omega_{k_{0}^{\prime}}} = \frac{1}{2} r_{e}^{2} \left(\frac{\omega_{0}^{\prime}}{\omega_{0}} \right)^{2} 
\left[ 1 + \cos^{2} \Theta_{0} + \left(\frac{\omega_{0}}{m} \right)^{2} \frac{\omega_{0}^{\prime}}{\omega_{0}} \left( 1 - \cos\Theta_{0} \right)^{2}  \right] \, ,
\label{eqb-5}
\end{eqnarray}
It is needless to say that one obtains the Thomson cross section by taking the limit $\omega_{0}/m \ll 1$ and $\omega_{0}^{\prime}/\omega_{0} \rightarrow 1$ in Eq.~(\ref{eqb-5}).

Now let us introduce the transformation from the electron rest frame to the CMB frame, where the electron is moving with a velocity $\beta$.  The photon energies $\omega$ and $\omega^{\prime}$ in the CMB frame are related to $\omega_{0}$ and $\omega_{0}^{\prime}$ by the Lorentz transformation
\begin{eqnarray}
&&\hspace{-10mm}
\omega = \omega_{0} \gamma \left( 1 - \beta \mu_{0} \right)\, ,
\label{eqb-6}  \\
&&\hspace{-10mm}
\omega^{\prime} = \omega_{0}^{\prime} \gamma \left( 1 - \beta \mu_{0}^{\prime} \right) \, ,
\label{eqb-7}
\end{eqnarray}
where $\mu_{0}=\cos \theta_{0}$ and $\mu_{0}^{\prime}=\cos \theta_{0}^{\prime}$.  With the variables $\omega$ and $\omega^{\prime}$ one obtains
\begin{eqnarray}
&&\hspace{-10mm}
\frac{d \sigma}{d \Omega_{k_{0}^{\prime}}} = \frac{1}{2} r_{e}^{2} \left(\frac{1-\beta\mu_0}{1-\beta\mu_0^{\prime}}\right)^2 \left(\frac{\omega^{\prime}}{\omega} \right)^{2} 
\left[ 1 + \cos^{2} \Theta_{0} + \left(\frac{\omega}{\gamma m} \right)^{2} \frac{\omega^{\prime}}{\omega} \frac{\left( 1 - \cos\Theta_{0} \right)^{2}}{(1-\beta \mu_{0})(1-\beta \mu_{0}^{\prime})}  \right] \, .
\label{eqb-8}
\end{eqnarray}
As seen from Eq.~(\ref{eqb-8}) the square bracket in the RHS is identical to $\bar{X}_{A} + \bar{X}_{B}$, where they are defined by Eqs.~(\ref{eq2-21}) and (\ref{eq2-22}).  Note that Eq.~(\ref{eqb-8}) is the expression in the hybrid coordinate system, where the energies are described in the CMB system, whereas the zenith angles are described in the electron rest frame.

  The cross section is defined by the transition rate divided by the flux of the incident particles.  The flux in the CMB frame is
\begin{eqnarray}
&&\hspace{-10mm}
j_{inc} \equiv \frac{p \cdot k}{E \omega} = 1 - \beta \mu  \, .
\label{eqb-9}
\end{eqnarray}
Therefore, one can write Eq.~(\ref{eq2-1}) in terms of the cross section in the CMB frame as follows:
\begin{eqnarray}
&&\hspace{-10mm}
\frac{\partial n(\omega)}{\partial t}  =  -2 \int \frac{d^{3}p}{(2\pi)^{3}}
 (1 - \beta \mu) \left( \frac{d \sigma}{d \Omega_{k^{\prime}}} \right) d \Omega_{k^{\prime}}  
\nonumber  \\
&&\hspace{+20mm}
\left\{ n(\omega)[1 + n(\omega^{\prime})] f(E) - n(\omega^{\prime})[1 + n(\omega)] f(E^{\prime}) \right\} \, .
\label{eqb-10}
\end{eqnarray}
Since the cross section is Lorentz invariant, one can rewrite Eq.~(\ref{eqb-10}) with the Klein-Nishina cross section in the hybrid system of Eq.~(\ref{eqb-8}) as follows:
\begin{eqnarray}
&&\hspace{-10mm}
\frac{\partial n(\omega)}{\partial t}  =  -2 \int \frac{d^{3}p}{(2\pi)^{3}}
 (1 - \beta \mu) \left( \frac{d \sigma}{d \Omega_{k_{0}^{\prime}}} \right) d \Omega_{k_{0}^{\prime}}  
\nonumber  \\
&&\hspace{+20mm}
\left\{ n(\omega)[1 + n(\omega^{\prime})] f(E) - n(\omega^{\prime})[1 + n(\omega)] f(E^{\prime}) \right\} \, .
\label{eqb-11}
\end{eqnarray}
Rewriting the phase space volume $d^{3}p$ by
\begin{eqnarray}
&&\hspace{-10mm}
d^3p = \frac{1}{\gamma^2(1-\beta\mu_0)^2}d^3p_0
\label{eqb-12}
\end{eqnarray}
and inserting Eqs.~(\ref{eqb-8}) and (\ref{eqb-12}), one finally obtains
\begin{eqnarray}
&&\hspace{-10mm}
\frac{\partial n(\omega)}{\partial \tau}
 = -\frac{3}{64\pi^2}
\int{d^3p_0}\int d\Omega_{k^{\prime}0}\frac{1}{\gamma^4}
\frac{1}{1-\beta\mu_0}
\frac{1}{(1-\beta\mu_0^{\prime})^2}
\left(\frac{\omega^{\prime}}{\omega}\right)^2\bar{X}
\nonumber \\
&&\hspace{+10mm}
\times
\left\{n(\omega)[1+n(\omega^{\prime})]p_e(E)
-n(\omega^{\prime})[1+n(\omega)]p_e(E^\prime)\right\}  \, .
\label{eqb-13}
\end{eqnarray}
In deriving Eq.~(\ref{eqb-13}) we used the relations $\gamma^2(1-\beta\mu) = (1-\beta\mu_0)^{-1}$, $f(E)=\pi^{2}n_{e}p_{e}(E)$, $d\tau=n_{e}\sigma_{T}dt$ and $\sigma_{T}=8\pi/3r_{e}^{2}$.  One finds that Eq.~(\ref{eqb-13}) is identical to Eq.~(\ref{eq2-27}).


\bibliography{apssamp}

\begin{thebibliography}{0}

\bibitem{suny72}
R. A. Sunyaev and Ya. B. Zeldovich, Comments Astrophys. Space Sci., {\bf 4}, 173 (1972).
\bibitem{suny80a}
R. A. Sunyaev and Ya. B. Zeldovich, Annu. Rev. Astron. Astrophys., {\bf 18}, 537 (1980).
\bibitem{suny80b}
R. A. Sunyaev and Ya. B. Zeldovich, Mon. Not. R. Astron. Soc., {\bf 190}, 413 (1980).
\bibitem{suny81}
R. A. Sunyaev and Ya. B. Zeldovich, Astrophys. Space Phys. Rev., {\bf 1}, 1 (1981).
\bibitem{birk99}
M. Birkinshaw, Physics Reports, {\bf 310}, 97 (1999).
\bibitem{carl02}
J. E. Carlstrom, G. P. Holder, and E. D. Reese, Annu. Rev. Astron. Astrophys., {\bf 40}, 643 (2002).
\bibitem{komp57}
A. S. Kompaneets, Soviet Physics JETP, {\bf 4}, 730 (1957).
\bibitem{tuck98}
W. Tucker, P. Blanco, S. Rappoport, L. David, D. Fabricant, E. E. Falco, W. Forman, A. Dressler and M. Ramella, Astrophys. J., {\bf 496}, L5 (1998).
\bibitem{alle02}
S. W. Allen, R. W. Schmidt, and A. C. Fabian, Mon. Not. R. Astron. Soc., {\bf 335}, 256 (2002).
\bibitem{wrig79}
E. L. Wright, Astrophys. J., {\bf 232}, 348 (1979).
\bibitem{reph95}
Y. Rephaeli, Astrophys. J., {\bf 445}, 33 (1995).
\bibitem{reph97}
Y. Rephaeli and D. Yankovitch, Astrophys. J., {\bf 481}, L55 (1997).
\bibitem{steb97}
A. Stebbins, preprint [astro-ph/9705178] (1997).
\bibitem{chal98}
A. Challinor and A. Lasenby, Astrophys. J., {\bf 499}, 1 (1998).
\bibitem{itoh98}
N. Itoh, Y. Kohyama and S. Nozawa, Astrophys. J., {\bf 502}, 7 (1998).
\bibitem{noza98}
S. Nozawa, N. Itoh and Y. Kohyama, Astrophys. J., {\bf 508}, 17 (1998).
\bibitem{chal99}
A. Challinor and A. Lasenby, Astrophys. J., {\bf 510}, 930 (1999).
\bibitem{sazo98a}
S. Y. Sazonov and R. A. Sunyaev, Astrophys. J., {\bf 508}, 1 (1998).
\bibitem{sazo98b}
S. Y. Sazonov and R. A. Sunyaev, Astronomy Letters {\bf 24}, 553 (1998).
\bibitem{itoh00}
N. Itoh, S. Nozawa and Y. Kohyama, Astrophys. J., {\bf 533}, 588 (2000).
\bibitem{itoh01}
N. Itoh, Y. Kawana, S. Nozawa and Y. Kohyama, Mon. Not. R. Astron. Soc., {\bf 327}, 567 (2001).
\bibitem{noza05}
S. Nozawa, N. Itoh and Y. Kohyama, Astron. Astrophys., {\bf 440}, 39 (2005).
\bibitem{noza00}
S. Nozawa, N. Itoh, Y. Kawana and Y. Kohyama, Astrophys. J., {\bf 536}, 31 (2000).
\bibitem{itoh02}
N. Itoh, T. Sakamoto, S. Kusano, Y. Kawana and S. Nozawa, Astron. Astrophys., {\bf 382}, 722 (2002).
\bibitem{itoh04}
N. Itoh and S. Nozawa, Astron. Astrophys., {\bf 417}, 827 (2004).
\bibitem{noza06}
S. Nozawa, N. Itoh, Y. Suda and Y. Ohata, IL Nuovo Cimento, {\bf 121B}, 487 (2006).
\bibitem{hans04}
S. H. Hansen, New Astron. {\bf 9}, 279 (2004).
\bibitem{enss00}
T. A. En{\ss}lin and C. R. Kaiser, Astron. Astrophys., {\bf 360}, 417 (2000).
\bibitem{cola03}
S. Colafrancesco, P. Marchegiani and E. Palladino, Astron. Astrophys., {\bf 397}, 27 (2003).
\bibitem{boeh08}
C. Boehm and J. Lavalle, arXiv:0812.3282 [astro-ph] (2008).
\bibitem{shim04}
M. Shimon and Y. Rephaeli, New Astronomy, {\bf 9}, 69 (2004).
\bibitem{chan50}
S. Chandrasekhar, {\it Radiative transfer}, Oxford, Clarendon Press (1950).
\bibitem{dolg01}
A. D. Dolgov, S. H. Hansen, S. Pastor and D. V. Semikoz, Astrophys. J., {\bf 554}, 74 (2001).
\bibitem{noza09}
S. Nozawa, Y. Kohyama and N. Itoh, in preparation.

\end{thebibliography}

\end{document}